\documentclass[twocolumn,apj]{openjournal}

\usepackage{xcolor}
\usepackage{textgreek}
\usepackage[utf8]{inputenc}
\usepackage[english]{babel}
\usepackage{amsmath}
\usepackage{hyperref}
\hypersetup{
    unicode, 
    colorlinks=true,
    linkcolor=linkcolor,
    citecolor=linkcolor,
    filecolor=linkcolor,
    urlcolor=linkcolor,
}
\usepackage{color,colortbl}
\definecolor{linkcolor}{rgb}{0.0,0.3,0.5}
\usepackage{tensind}
\DeclareGraphicsExtensions{.bmp,.png,.jpg,.pdf}
\usepackage{verbatim}
\usepackage[normalem]{ulem}
\usepackage{orcidlink}
\usepackage{soul}
\newcommand{\orcidauthor}[3]{\author{\href{http://orcid.org/#1}{#2$^{#3}$}}}

\graphicspath{ {./} }

\begin{document}
\title{\vspace{-0.8cm}N\lowercase{eutrino} C\lowercase{onstraints} \lowercase{on} B\lowercase{lack} H\lowercase{ole} F\lowercase{ormation} \lowercase{in} M31\vspace{-1.5cm}}

\orcidauthor{0000-0002-7443-2215}{Yudai Suwa}{1,2 *}
\orcidauthor{0000-0002-9234-813X}{Ryuichiro Akaho}{3}
\orcidauthor{0000-0003-4136-2086}{Yosuke Ashida}{4}
\orcidauthor{0000-0003-1409-0695}{Akira Harada}{5,6}
\orcidauthor{0000-0003-3273-946X}{Masayuki Harada}{7}
\orcidauthor{0000-0003-0437-8505}{Yusuke Koshio}{8,9}
\orcidauthor{0000-0002-0827-9152}{Masamitsu Mori}{10,11}
\orcidauthor{0000-0003-4408-6929}{Fumi Nakanishi}{8}
\orcidauthor{0000-0001-6330-1685}{Ken'ichiro Nakazato}{12}
\orcidauthor{0000-0002-9224-9449}{Kohsuke Sumiyoshi}{11}
\orcidauthor{0000-0002-0969-4681}{Roger A. Wendell}{13,9}
\orcidauthor{0000-0001-7305-1683}{Masamichi Zaizen}{1}

\affiliation{$^{1}$Department of Earth Science and Astronomy, The University of Tokyo, Tokyo 153-8902, Japan}
\affiliation{$^{2}$Center for Gravitational Physics and Quantum Information, Yukawa Institute for Theoretical Physics, Kyoto University, Kyoto 606-8502, Japan}
\affiliation{$^{3}$Faculty of Science and Engineering, Waseda University, Tokyo 169-8555, Japan}
\affiliation{$^{4}$Department of Physics, Tohoku University, Sendai, Miyagi 980-8578, Japan}
\affiliation{$^{5}$National Institute of Technology, Ibaraki College, Hitachinaka, Ibaraki 312-8508, Japan}
\affiliation{$^{6}$Interdisciplinary Theoretical and Mathematical Sciences Program (iTHEMS), RIKEN, Wako, Saitama 351-0198, Japan}
\affiliation{$^{7}$Kamioka Observatory, Institute for Cosmic Ray Research, The University of Tokyo, Gifu 506-1205, Japan}
\affiliation{$^{8}$Department of Physics, Okayama University, Okayama 700-8530, Japan}
\affiliation{$^{9}$Kavli Institute for the Physics and Mathematics of the Universe (Kavli IPMU, WPI), Todai Institutes for Advanced Study, The University of Tokyo, Kashiwa, Chiba 277-8583, Japan}
\affiliation{$^{10}$Division of Science, National Astronomical Observatory of Japan, 2-21-1 Osawa, Mitaka, Tokyo 181-8588, Japan}
\affiliation{$^{11}$National Institute of Technology, Numazu College, Numazu, Shizuoka 410-8501, Japan}
\affiliation{$^{12}$Faculty of Arts and Science, Kyushu University, Fukuoka 819-0395, Japan}
\affiliation{$^{13}$Department of Physics, Kyoto University, Kyoto 606-8502, Japan}

\email{* suwa@yukawa.kyoto-u.ac.jp}

\begin{abstract}
We investigate neutrino signals associated with black hole formation resulting from the gravitational collapse of massive stars, motivated by the candidate failed supernova M31-2014-DS1 in the Andromeda Galaxy (M31). By compiling numerical simulation results for stellar collapse, we predict the expected neutrino emission and compare these predictions with observational limits from Super-Kamiokande (SK). The simulations reveal a characteristic precursor signal consisting of a short, intense burst whose average neutrino energy rises rapidly and then ceases abruptly once the black hole forms. We examine several nuclear equations of state, specifically the Lattimer \& Swesty, Shen, Togashi, and SFHo models, to evaluate how the emission depends on neutron-star properties and nuclear-physics uncertainties. Comparison of the predicted event counts with SK’s non-detection of neutrinos coincident with M31-2014-DS1 already rules out part of the model space and highlights the sensitivity of current neutrino detectors to both progenitor mass and the EOS. These findings demonstrate the capability of neutrino astronomy to probe core collapse and black hole formation in failed supernova scenarios.
\end{abstract}


\maketitle

\section{Introduction}
\label{sec:intro}

Recently, the disappearance of a massive star, identified as M31-2014-DS1, in the Andromeda galaxy (M31) was reported without a typical luminous supernova explosion \citep{2024arXiv241014778D}. Such phenomena, often referred to as ``failed supernovae,'' involve black hole formation via gravitational collapse without a significant optical outburst. Theoretical studies have long predicted that certain massive stars could end their evolution through a quiet collapse directly into black holes, particularly stars characterized by highly compact cores or masses exceeding specific thresholds \citep{2011ApJ...730...70O, 2016ApJ...821...38S}. Systematic numerical simulations by \citet{2011ApJ...730...70O} demonstrated a strong correlation between core compactness and the likelihood of failed explosions, emphasizing that stars exceeding a critical value of the compactness parameter are likely to collapse silently into black holes rather than undergoing successful explosions.

Despite numerous numerical simulations and theoretical predictions, confirming failed supernovae observationally remains difficult because their electromagnetic output is intrinsically faint. Only a handful of credible candidates have been reported to date. The first compelling case, N6946-BH1 in NGC~6946, exhibited a modest optical outburst before disappearing \citep{2015MNRAS.450.3289G,2017MNRAS.468.4968A}, and multi-wavelength follow-up favours its direct collapse into a black hole \citep{2017MNRAS.469.1445A} (but see \citealt{2024ApJ...964..171B}). Systematic archival \textit{Hubble Space Telescope} monitoring has since revealed at least one additional disappearance of a $25$-$30\,M_{\odot}$ yellow supergiant \citep{2015MNRAS.453.2885R}, while the apparent loss of a luminous-blue-variable-like star in the extremely metal-poor dwarf galaxy PHL~293B offers another promising example \citep{2020MNRAS.496.1902A}. Nevertheless, distinguishing genuine failed supernovae from impostors such as luminous-blue-variable eruptions or stellar mergers remains challenging. The recently reported M31-2014-DS1, only $770$\,kpc away, therefore provides a uniquely accessible opportunity to investigate this elusive phenomenon in greater detail than previous events.

Neutrinos emitted during stellar gravitational collapse serve as unique diagnostic probes, as they escape from the dense stellar core with weak interaction with the surrounding matter \citep{2012ARNPS..62...81S,2017hsn..book.1575J,2018JPhG...45d3002H,2019ARNPS..69..253M}. Numerical simulations predict distinct neutrino signatures associated with failed supernovae, notably a sharp cutoff in neutrino emissions coinciding with the formation of the black hole horizon, along with potential short-duration energetic neutrino bursts just before collapse \citep{2004ApJS..150..263L,2006PhRvL..97i1101S}. For instance, simulations by \citet{2020PhRvD.101l3013W} suggest that failed supernovae could produce neutrino fluxes comparable to or exceeding those of successful explosions, albeit significantly shorter in duration. Detailed numerical modelling also highlights complex neutrino emission patterns, potentially including oscillatory features caused by dynamic instabilities such as the standing accretion shock instability, providing a valuable diagnostic signature for collapse dynamics. By contrast, scenarios in which the shock is successfully revived yet late-time fallback accretion eventually drives the protoneutron star to collapse into a black hole can yield even larger total neutrino energies; however, these cases should be accompanied by a bright electromagnetic display, and can therefore be ruled out when no such counterpart is observed.

From 2008 to 2018, the Super-Kamiokande (SK) neutrino observatory conducted nearly a decade of continuous neutrino monitoring, actively searching for astrophysical neutrino bursts associated with gravitational collapse events \citep{2022ApJ...938...35M}. Although no significant neutrino signals coinciding with the timing of M31-2014-DS1 were detected, these null detections still provide important constraints on theoretical models of neutrino emissions and core-collapse mechanisms. Present neutrino detectors, including SK, primarily possess sensitivity limited to events within or near our galaxy, typically within hundreds of kpc. Therefore, while the non-detection from M31 is consistent with current detector capabilities, SK could have had a 20\%-40\% chance of detection at the distance of M31 assuming the neutrino emission model of \citet{2010PhRvD..81h3009N}. This emphasizes the necessity for future detectors, such as Hyper-Kamiokande, JUNO, and DUNE, designed to improve detection sensitivity to extragalactic core-collapse neutrinos. Upcoming detectors may feasibly detect neutrino emissions from events such as the failed supernova candidate M31-2014-DS1, significantly advancing our ability to study stellar collapse physics and the properties of black hole formation.

In this paper, we systematically survey theoretical simulations predicting neutrino emissions from black-hole-forming stellar collapses and compare these predictions rigorously against observational constraints from SK's neutrino monitoring data. Through this comparative analysis, we aim to refine constraints on the mechanisms and conditions of failed supernovae. These refinements will provide clearer insights into the late evolutionary stages of massive stars and the physical conditions that facilitate their silent collapse into black holes.
In Section \ref{sec:simulation}, we discuss the characteristic neutrino signatures predicted by black hole formation simulations. Section \ref{sec:constraints} evaluates the expected neutrino event numbers and observational constraints derived from SK data. In Section \ref{sec:prospects}, we discuss future detection prospects in the upcoming Hyper-Kamiokande era. Finally, Section \ref{sec:summsary} provides a concise summary of our findings and their implications.

\section{Neutrino Signatures Predicted by Black Hole Formation Simulations}
\label{sec:simulation}

    \setlength{\tabcolsep}{3pt} 
    \renewcommand{\arraystretch}{1.1} 
    \setlength\extrarowheight{4pt}
    \begin{table*}[tbp]
        \centering
        \caption{Summary of Simulations}
        \label{table:table}
        \begin{tabular}{ llllccccccc }  
        \hline
            References\footnote{In this study, we only use simulation results that include comprehensive neutrino data.} & Progenitor mass & EOS & Gravity  & $t_{\rm BH}$\footnote{Postbounce time.} & $E_{\nu_e}$ & $E_{\bar\nu_e}$ & $E_{\nu_X}$ & $\left<\varepsilon_{\nu_e}\right>$ & $\left<\varepsilon_{\bar\nu_e}\right>$ & $\left<\varepsilon_{\nu_X}\right>$\\
            & ($M_\odot$) & & & (s) & ($10^{53}$ erg) & ($10^{53}$ erg) & ($10^{53}$ erg) & (MeV) & (MeV) & (MeV)\\
            \hline \hline     
            \cite{2007ApJ...667..382S} & 40 (WW95\footnote{\cite{1995ApJS..101..181W}.}) & LS180      & fGR\footnote{Fully GR.}       & 0.56  & 0.554 & 0.467 & 0.228 & 16.3 & 19.5 & 21.5\\
            \cite{2007ApJ...667..382S} & 40 (WW95)                                       & Shen-TM1   & fGR                           & 1.34  & 1.46  & 1.35  & 0.526 & 20.3 & 23.2 & 23.9\\
            \cite{2008ApJ...688.1176S} & 40 (H95\footnote{\cite{1995PThPh..94..663H}.})  & LS180      & fGR                          & 0.36  & 0.334 & 0.271 & 0.160 & 13.5 & 16.8 & 21.9\\
            \cite{2008ApJ...688.1176S} & 50 (TUN07\footnote{\cite{2007ApJ...660..516T}.})& Shen-TM1   & fGR                          & 1.51  & 1.35  & 1.27  & 0.526 & 20.0 & 23.1 & 24.2\\
            \cite{2008ApJ...688.1176S} & 50 (TUN07)                                      & LS180      & fGR                           & 0.507 & 0.450 & 0.372 & 0.191 & 15.7 & 19.0 & 21.2\\
            \cite{fisc09}\footnote{See also \cite{2004ApJS..150..263L}.} & 40 (WW95)     & LS180      & fGR                          & 0.435 & 0.507 & 0.376 & 0.231 & 14.1 & 14.6 & 19.7\\
            \cite{fisc09}              & 40 (WW95)                                       & Shen-TM1   & fGR                           & 1.40  & 1.73  & 1.53  & 0.715 & 16.0 & 18.4 & 21.0\\
            \cite{2010PhRvD..81h3009N} & 40 (WW95)                                       & LS220      & fGR                           & 0.780 & 0.729 & 0.627 & 0.382 & 17.3 & 20.1 & 24.1\\
            \cite{2014PhDT.......436H} & 40 (WW95)                                       & LS180      & eGR\footnote{Effectively GR.} & 0.435 & 0.422 & 0.337 & 0.209 & 13.8 & 17.1 & 18.3\\
            \cite{2014PhDT.......436H} & 40 (WW95)                                       & LS220      & eGR                           & 0.55  & 0.525 & 0.436 & 0.279 & 14.4 & 17.7 & 19.2\\
            \cite{2014PhDT.......436H} & 25 (WHW02\footnote{\cite{2002RvMP...74.1015W}.})& LS220      & eGR                          & 1.225 & 0.696 & 0.632 & 0.331 & 15.3 & 18.5 & 17.7\\
            \cite{2014PhDT.......436H} & 40 (WHW02)                                      & LS220      & eGR                           & 1.93  & 0.852 & 0.796 & 0.402 & 15.8 & 18.8 & 17.4\\
            \cite{2019ApJ...887..110S} & 50 (TUN07)                                      & Shen-TM1e  & fGR                           & 1.15  & 0.941 & 0.850 & 0.330 & 18.7 & 21.9 & 21.6\\
            \cite{2019ApJ...887..110S} & 40 (WW95)                                       & Shen-TM1e  & fGR                           & 1.103 & 1.15  & 1.05  & 0.422 & 19.3 & 22.4 & 23.1\\
            \cite{2020PhRvD.101l3013W} & 40 (WH07\footnote{\cite{2007PhR...442..269W}.}) & LS220      & eGR                          & 0.57  & 0.572 & 0.539 & 0.375 & 16.2 & 18.8 & 20.2\\
            \cite{2021PASJ...73..639N} & 30 (N13\footnote{\cite{2013ApJS..205....2N}.})  & LS220      & fGR                          & 0.342 & 0.403 & 0.287 & 0.211 & 12.5 & 16.4 & 22.3\\
            \cite{2021PASJ...73..639N} & 30 (N13)                                        & Togashi    & fGR                           & 0.533 & 0.685 & 0.533 & 0.289 & 16.1 & 20.4 & 23.4\\
            \cite{2021PASJ...73..639N} & 30 (N13)                                        & Shen-TM1   & fGR                           & 0.842 & 0.949 & 0.81  & 0.400 & 17.5 & 21.7 & 23.4\\
            \cite{2021ApJ...909..169K} & 40 (WW95)                                       & LS220      & eGR                           & 0.57  & 0.938 & 0.862 & 0.483 & 15.7 & 18.7 & 17.6\\
            \cite{2021ApJ...909..169K} & 40 (WHW02)                                      & LS220      & eGR                           & 2.11  & 0.544 & 0.449 & 0.281 & 14.4 & 17.6 & 18.8\\
            \cite{2024ApJ...964..143K} & 40 (WHW02)                                      & DD2F RDF-1.7 & fGR                         & 1.03  & 0.368 & 0.384 & 0.132 & 13.3 & 15.6 & 14.2\\
            \cite{2025arXiv250307531C}\footnote{See also \cite{2024arXiv241207831B}, which present three-dimensional simulations.} & 12.25 (S16\&18\footnote{\cite{2016ApJ...821...38S,2018ApJ...860...93S}.}) & SFHo    & eGR  &  $>$2.09 & 0.563 & 0.511 & 0.297 & 13.9 & 16.3 & 15.5\\
            \cite{2025arXiv250307531C} & 14 (S16\&18)                                    & SFHo       & eGR                           & $>$2.82 & 0.768 & 0.711 & 0.393 & 15.0 & 17.3 & 15.9\\
            \cite{2025arXiv250307531C} & 19.56 (S16\&18)                                 & SFHo       & eGR                           & 3.89  & 0.906 & 0.889 & 0.694 & 12.9 & 15.4 & 16.0\\
            \cite{2025arXiv250307531C} & 23 (S16\&18)                                    & SFHo       & eGR                           & 6.23  & 0.776 & 0.736 & 0.609 & 12.4 & 14.7 & 14.8\\
            \cite{2025arXiv250307531C} & 40 (S16\&18)                                    & SFHo       & eGR                           & 1.76  & 0.798 & 0.766 & 0.499 & 13.4 & 15.8 & 16.0\\
            \cite{2025arXiv250307531C} & 100 (S16\&18)                                   & SFHo       & eGR                           & 0.44  & 0.529 & 0.462 & 0.246 & 12.9 & 15.1 & 17.1\\
            unpublished\footnote{The same code as \cite{2010PhRvD..81h3009N}.} & 40 (WW95) & Togashi  & fGR                  & 0.927 & 0.824 & 0.705 & 0.471 & 18.1 & 20.7 & 25.7\\
            \hline \hline  
        \end{tabular}
    \end{table*}

In this section, we provide a detailed summary of neutrino emission characteristics predicted by numerical simulations of black hole formation in failed supernovae. Theoretical modeling, particularly using general-relativistic simulations, has significantly advanced our knowledge of neutrino signals associated with these extreme astrophysical events.

General-relativistic (GR) simulations consistently predict a distinct neutrino emission pattern during gravitational collapse that leads to black hole formation \citep{2004ApJS..150..263L,2007ApJ...667..382S,fisc09}. After core bounce, a protoneutron star (PNS) forms and emits intense neutrino radiation as it accretes infalling stellar material. In the case of failed supernovae, the neutrino luminosity can reach values on the order of $10^{53}$ erg s$^{-1}$ per neutrino flavor. However, this luminosity is sustained only briefly ($\sim$0.5-1 s) compared with typical successful supernova events. As the PNS nears gravitational instability and subsequent black hole formation, neutrino spectra undergo notable hardening, marked by a rapid increase in mean neutrino energy. Multidimensional simulations also predict a final energetic neutrino burst emitted just before the disappearance of the neutrino source into the forming black hole \citep{2023MNRAS.526..152K}. Immediately after the event horizon forms, neutrino emission abruptly ceases, producing a clear and identifiable signature--a sudden cutoff in the neutrino flux \citep[but see also][]{2022ApJ...926..212G}.

Detecting these characteristic neutrino signals from failed supernovae would offer a direct observational confirmation of black hole formation processes in massive stellar collapses. Such neutrino signals may represent solely observational evidence of these otherwise electromagnetically dark events. In the following, we present a comprehensive overview of simulations performed to date, summarizing their neutrino emission predictions, including luminosities, spectral properties, and temporal features. This summary provides a foundation for subsequent comparison against observational limits and evaluating theoretical predictions for black hole formation in stellar collapses.

Table \ref{table:table} gives a summary of simulation results for core-collapse supernovae resulting in black hole formation from various studies. Columns show the references, progenitor mass, the employed equation of state (EOS), treatment of gravity (fGR for fully general relativistic and eGR for effectively general relativistic; the latter employs Newtonian hydrodynamics combined with an approximate gravitational potential incorporating general relativistic effects), black hole formation time from core bounce ($t_\mathrm{BH}$), emitted neutrino energies ($E_{\nu_e}$, $E_{\bar{\nu}_e}$, $E_{\nu_X}$), and corresponding mean neutrino energies ($\langle \varepsilon_{\nu_e}\rangle$, $\langle \varepsilon_{\bar{\nu}e}\rangle$, $\langle\varepsilon_{\nu_X}\rangle$).
Simulation outcomes reveal a systematic dependence of the neutrino signal on the maximum mass supported by the EOS with the finite temperature correction.\footnote{In this paper we employ the entropy $s=$ 3 or 4$\,k_{\mathrm{B}}$/baryon for the finite temperature correction.} EOSs with larger maximum masses sustain longer emission phases, yielding larger integrated neutrino energies and correspondingly higher expected event counts. In contrast, EOSs with smaller maximum masses terminate the emission sooner, producing smaller total energies and fewer events \citep{2006PhRvL..97i1101S}. Although progenitor mass modulates these trends -- more massive stars generally enhance the overall neutrino output -- the distinction between EOSs with larger and smaller maximum masses remains the primary factor governing both the duration and the amplitude of the predicted neutrino signal.

The EOS significantly affects neutrino emission by determining the thermodynamic and nuclear properties of dense matter during collapse. Several EOS models have been employed in the simulations, which are summarized here. 
The Lattimer-Swesty EOS \citep{1991NuPhA.535..331L}, based on a Skyrme-type nuclear interaction and a compressible liquid-drop model, has been widely used in earlier simulations. This EOS includes variations characterized by different nuclear incompressibility parameters ($K$); among the simulations presented in this work, only the $K=180$\,MeV (LS180) and $K=220$\,MeV (LS220) variants are used. 
The Shen EOS family \citep{1998NuPhA.637..435S,2020ApJ...891..148S}, constructed within relativistic mean-field theory, is represented here by two parameter sets: the original one (Shen-TM1) and the updated one (Shen-TM1e). Shen-TM1e adopts a reduced symmetry-energy slope compared with Shen-TM1, resulting in a slightly softer pressure at supra-nuclear densities and hence altered neutrino luminosities and spectra. 
The Togashi EOS \citep{2017NuPhA.961...78T} employs advanced nuclear many-body calculations, offering detailed composition and structure predictions at high densities. 
The DD2F RDF EOS \citep{2021PhRvD.103b3001B} is a hadron-quark matter hybrid EOS with a first-order phase transition, and we employ the DD2F RDF-1.7 variant, where 1.7 denotes a specific parameter set in this EOS series characterizing the strength of quark-matter interactions.
Lastly, the SFHo EOS \citep{2013ApJ...774...17S}, derived from relativistic mean-field models constrained by experimental nuclear data, is frequently utilised owing to its realistic treatment of nuclear-matter properties at densities relevant to PNS formation and neutrino emission.
Since temperatures during black hole formation are very high, it is essential to consider finite-temperature effects rather than zero-temperature properties when comparing EOS models. In this analysis, we adopt an entropy value of $4\,k_\mathrm{B}$/baryon as a typical reference condition to evaluate each EOS consistently under relevant collapse conditions. The corresponding maximum gravitational masses of PNSs, $M_{s=4}^{\rm max}$, predicted by each EOS are approximately: LS180 (2.0~$M_{\odot}$), LS220 (2.1~$M_{\odot}$), SFHo (2.3~$M_{\odot}$), Togashi (2.4~$M_{\odot}$), Shen-TM1e (2.5~$M_{\odot}$) and Shen-TM1 (2.6~$M_{\odot}$) \citep[see, e.g.,][]{2013ApJ...774...17S,2020ApJ...894....4D}. For reference, the DD2F RDF-1.7 EOS yields a similar maximum mass of about 2.1~$M_\odot$, where the value at $3\,k_{\mathrm{B}}$/baryon is adopted based on the value reported in the literature.
These results indicate that EOSs predicting larger maximum masses also yield longer-lived PNSs. Consequently, such EOSs tend to produce longer neutrino-emission phases and higher total neutrino event counts, whereas those with smaller maximum masses result in earlier emission termination and smaller total counts. The instantaneous luminosity, however, does not necessarily increase with maximum mass, as shown, for example, by \citet{2006PhRvL..97i1101S}.
Comparing simulation results across these EOS models helps constrain uncertainties in nuclear physics that affect neutrino emission modeling from stellar collapses.

\section{Expected Neutrino Events and Constraints from Super-Kamiokande}
\label{sec:constraints}

\begin{center}
    \begin{figure*}[t]
        \centering
    	\includegraphics[width=0.7\textwidth]{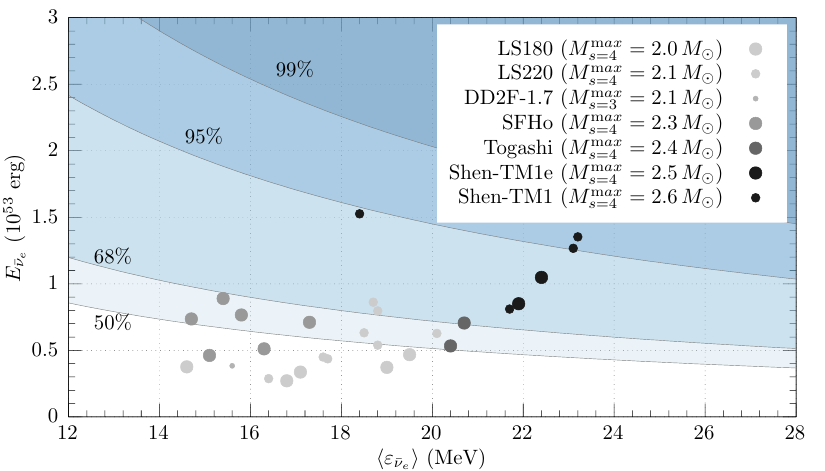}
    	\caption{Total emitted electron antineutrino energy $E_{\bar{\nu}_e}$ versus mean energy $\langle\varepsilon_{\bar{\nu}_e}\rangle$ predicted by black-hole-forming core-collapse simulations. The shaded bands indicate regions that would have produced $\ge 2$ correlated events in SK with probabilities of 50\,\%, 68\,\%, 95\,\%, and 99\,\% (light to dark blue), assuming Poisson statistics for a source at the distance of M31. Filled circles show individual simulation results for seven nuclear equations of state (EOS): LS180, LS220, DD2F RDF-1.7, SFHo, Togashi, Shen-TM1e, and Shen-TM1. Both grey tone and symbol size encode the maximum gravitational mass of a non-rotating protoneutron star supported by each EOS: lighter symbols correspond to softer EOSs with the maximum gravitational mass of protoneutron star with thermal correction, $M^{\max}_{s=4}=2.0\text{--}2.1\,M_{\odot}$ (LS180, LS220), whereas darker symbols mark EOSs with $M^{\max}_{s=4}=2.5\text{--}2.6\,M_{\odot}$ (Shen-TM1e, Shen-TM1). Models that lie outside the shaded regions, where fewer than two events would be expected, remain compatible with the SK non-detection, illustrating the strong dependence of the predicted neutrino signal on EOS stiffness and the associated maximum neutron-star mass.}
        \label{fig:fig}
    \end{figure*}
\end{center}

The expected number of neutrino events detected can be estimated by:
\begin{eqnarray}
    N_\nu&=&\frac{2}{18}\frac{M_{\rm det}}{m}\frac{E_\nu}{4\pi D^2\left<\varepsilon_\nu\right>}\left<\sigma\right>\nonumber\\
    &=&3.48
    \left(\frac{D}{770\,{\rm kpc}}\right)^{-2}
    \left(\frac{\left<\varepsilon_\nu\right>}{15\,{\rm MeV}}\right)
    \left(\frac{E_\nu}{10^{53}\,{\rm erg}}\right),
\end{eqnarray}
where $M_{\rm det}$ is the detector mass (22.5 kton, corresponding to the fiducial volume of SK), $m$ is the nucleon mass, $D$ is the distance between the supernova and Earth (770 kpc, corresponding to the distance to M31), and $\sigma$ is the inverse beta decay cross-section ($p+\bar{\nu}_e\to n+e^{+}$).\footnote{In this calculation, we focus exclusively on inverse beta decay (IBD) interactions. Although the observational limits derived by \citet{2022ApJ...938...35M} were obtained considering all neutrino interactions within the relevant energy window, contributions from channels other than IBD are relatively minor. Thus, limiting our estimate to IBD provides a reasonable approximation.} For the cross-section, we adopt $\sigma(\varepsilon_\nu)=\sigma_0(\varepsilon_\nu/{\rm MeV})^2$ with $\sigma_0=9.4\times 10^{-44}\,{\rm cm}^2$ \citep{1996slfp.book.....R} and assume a thermal neutrino spectrum (see Section 4 in \citealt{2021PTEP.2021a3E01S} for more details).\footnote{In this calculation, a purely thermal neutrino spectrum is assumed. However, a nonthermal neutrino component might also arise due to compressional flows onto the newly formed black hole or turbulent motion \citep{2013MNRAS.428.2443S,2019PTEP.2019h3E04S,2021MNRAS.502...89N}, potentially increasing the detectability significantly. Precise estimates should consider such effects, as well as detailed detector responses and efficiencies.} The factor $2/18$ accounts for the mass fraction of free protons in water, making $2M_{\rm det}/18m$ the total number of hydrogen nuclei within the detector volume.

Using ten years of SK‐IV data (2008--2018), \citet{2022ApJ...938...35M} searched for neutrino bursts from distant core-collapse supernovae with a statistical clustering algorithm. To suppress accidental coincidences, the analysis required either (i) at least two candidate events within a 0.5~s or 2~s window, or (ii) at least four events within a 10~s window. Despite a background rejection rate exceeding $99.3\,\%$ and a detection efficiency of roughly $74\,\%$, no neutrino cluster passed these criteria. In this analysis, although the nominal analysis threshold is 5.5 MeV, an effective energy cut of about 15 MeV is applied to suppress background events. Figure \ref{fig:fig}, described in the following, instead assumes an idealized, background-free detector, representing the ultimate sensitivity rather than the actual SK condition.

Figure \ref{fig:fig} shows the relationship between the total emitted energy by electron type antineutrinos $E_{\bar\nu_e}$ and the mean neutrino energy $\langle\varepsilon_{\bar\nu_e}\rangle$. The shaded regions represent excluded parameter space based on the non-detection of correlated neutrino events in SK, assuming Poisson statistics. Each contour corresponds to a confidence level of 50\%, 68\%, 95\%, or 99\%, and indicates the set of emission parameters ($E_{\bar\nu_e}$, $\langle\varepsilon_{\bar\nu_e}\rangle$) for which the probability of detecting two or more events would have exceeded the given confidence level. For example, an expected number of approximately 1.68 events corresponds to a 50\% chance of detecting two or more events, while about 2.35, 4.74, and 6.64 expected events correspond to the 68\%, 95\%, and 99\% confidence levels, respectively.

\begin{center}
    \begin{figure*}[t]
        \centering
    	\includegraphics[width=0.7\textwidth]{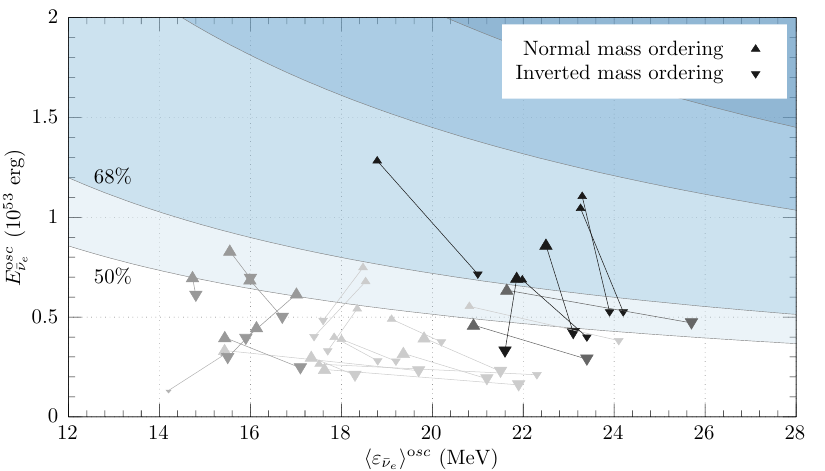}
    	\caption{The same as Figure \ref{fig:fig}, but with neutrino oscillation taken into account. Triangles indicate model predictions under normal (upward triangles) and inverted (downward triangles) neutrino mass orderings. 
}
        \label{fig:fig2}
    \end{figure*}
\end{center}

Additionally, we have considered the effects of neutrino flavor conversions by matter effect \citep{2000PhRvD..62c3007D} on the detectable neutrino signals at Earth. In the presence of Mikheyev-Smirnov-Wolfenstein (MSW) flavor conversion, the observed flux of electron antineutrinos can be expressed as a linear combination of the original electron and non-electron antineutrino fluxes:
\begin{equation}
F_{\bar{\nu}_e}^{\mathrm{osc}} =
\begin{cases}
\cos^2\theta_{12}\,F_{\bar{\nu}_e}^{\mathrm{src}} + \sin^2\theta_{12}\,F_{\bar{\nu}_X}^{\mathrm{src}} & \text{(normal ordering)} \\
F_{\bar{\nu}_X}^{\mathrm{src}} & \text{(inverted ordering)}
\end{cases}.
\end{equation}
Here, $\theta_{12}$ is the solar mixing angle, for which the current best-fit value is $\sin^2\theta_{12} \simeq 0.304$ \citep{2020JHEP...09..178E}, giving $\cos^2\theta_{12} \simeq 0.696$. This flavor transformation alters both the total observed energy and mean energy of electron antineutrinos, affecting their detectability in terrestrial detectors.
Figure \ref{fig:fig2} illustrates the relationship between total neutrino energy and mean neutrino energy under normal and inverted neutrino mass orderings. As shown, neutrino oscillations substantially alter both the total observed neutrino energy and mean neutrino energy, shifting theoretical predictions across the observational constraint boundaries. Notably, the mass ordering significantly influences detectability, highlighting the importance of accurately modeling neutrino flavor conversions when interpreting observations and constraining theoretical models.

Different EOS models -- LS180 ($M_{s=4}^{\rm max}=2.0,M_{\odot}$), LS220 ($M_{s=4}^{\rm max}=2.1,M_{\odot}$), DD2F RDF-1.7 ($M_{s=3}^{\rm max}=2.1,M_{\odot}$), SFHo ($M_{s=4}^{\rm max}=2.3,M_{\odot}$), Togashi ($M_{s=4}^{\rm max}=2.4,M_{\odot}$), Shen-TM1e  ($M_{s=4}^{\rm max}=2.5,M_{\odot}$) and Shen-TM1 ($M_{s=4}^{\rm max}=2.6,M_{\odot}$) -- are represented by symbols with varying shades and sizes. Most simulation results cluster around or below the 50\% confidence boundary, indicating good consistency with observational constraints, particularly at lower neutrino energies. The Shen-TM1 EOS, associated with the largest maximum PNS mass ($2.6 M_{\odot}$), predicts higher total neutrino energies, yet these results also fall within the observational limits. This clearly demonstrates how neutrino emission properties depend on the chosen EOS model.

\section{Prospects for the Hyper-Kamiokande Era}
\label{sec:prospects}

As discussed in the previous section, although we derived new limits for this event, the predicted number of events for many models fell sufficiently below these limits, preventing the imposition of more stringent constraints. Consequently, we explore the potential constraints achievable if Hyper-Kamiokande were operational during a future black hole formation event in a nearby galaxy.

Hyper-Kamiokande (HK), with its 187\,kton fiducial mass \citep{2018arXiv180504163H}, offers roughly eight times the target mass of SK. A model that would yield $N_\nu = 1.68$ events at SK (the 50\,\% Poisson contour) therefore gives \(N_{\mathrm{HK}} \simeq 14\,(770\,\mathrm{kpc}/D)^{2}\) in HK. This translates to about 14 events from M31 ($D \sim 770$\,kpc), 11 from M33 (860\,kpc), 15 from IC\,10 (750\,kpc), and 34 from NGC\,6822 (490\,kpc); a collapse in the Large Magellanic Cloud ($D \approx 50$\,kpc) would generate nearly $3\times10^{3}$ events, allowing millisecond-scale timing and detailed spectral analysis. Even the ten-to-few-dozen events expected from other Local Group galaxies will be sufficient to differentiate between competing EOS and probe the neutrino-mass ordering, converting current nondetections into definitive measurements and tightening constraints on black-hole-forming collapses across the Local Group.

\section{Summary}
\label{sec:summsary}

In this paper, we explored neutrino emission characteristics resulting from black hole formation in massive stellar collapses, motivated by the recent observation of the failed supernova candidate M31-2014-DS1 in the Andromeda Galaxy (M31). We systematically surveyed numerical simulations predicting neutrino signals associated with gravitational collapse leading to black holes. These simulations reveal characteristic neutrino signatures marked by intense bursts, spectral hardening, and abrupt termination at the moment of black hole formation.

We presented a detailed comparison of neutrino emission predictions across various equations of state (EOS), including Lattimer-Swesty (LS180, LS220), Shen (Shen-TM1e, Shen-TM1), Togashi, DD2F RDF, and SFHo. Results clearly illustrate the strong dependence of neutrino emission on EOS choice, with models employing stiff EOS and higher progenitor masses generally predicting more energetic neutrino outputs.

Using these theoretical predictions, we estimated the expected neutrino detection numbers at Super-Kamiokande (SK) for a collapse event occurring at the distance of M31 ($\sim$770 kpc). The observed non-detection of neutrino events around the time of M31-2014-DS1 places meaningful constraints on model parameters. Specifically, parameter combinations predicting higher neutrino energies and fluxes can be excluded at high confidence levels, thus providing valuable feedback to theoretical modeling efforts.
In this analysis, we did not consider the details of event selection such as energy cuts, in order to keep the estimate simple. When applying to real data, various cut conditions need to be taken into account, which would generally relax the constraints compared to those presented here \citep{2025arXiv251103470N}.

Neutrino observations provide a direct way to study stellar collapse and black-hole formation. The next generation of detectors--Hyper-Kamiokande, JUNO, and DUNE--will extend this power to extragalactic failed supernovae and tighten the limits on core-collapse models. To make the best use of these data, we also need steady theoretical progress: systematic simulations covering a wide range of progenitor stars and nuclear EOSs are necessary to build the large template bank of predicted neutrino signals that these experiments will compare with. By combining these observational and theoretical efforts, we will soon be able to see in new detail how massive stars end quietly and how stellar-mass black holes are born.

\section*{Acknowledgments}

We thank NK Largani and T. Fischer for providing their data.
This work is supported by JSPS KAKENHI grant Nos. JP21K13913, JP23KJ2150, JP24H02236, JP24H02245, JP24K00632, JP24K00668, JP24K07021  JP25K01035, and JP25K17399. 
Numerical computations were, in part, carried out on a computer cluster at CfCA of the National Astronomical Observatory of Japan. This work was supported in part by MEXT as ``Program for Promoting Research on the Supercomputer Fugaku'' (Toward a unified view of the universe: from large scale structures to planets) and, in part by the Inter-University Research Program of the Institute for Cosmic Ray Research (ICRR), the University of Tokyo.

\bibliographystyle{hapj}


\end{document}